\documentclass[12pt]{article}
\usepackage[T1]{fontenc}
\usepackage[utf8]{inputenc}
\usepackage[a4paper]{geometry}
\usepackage{xcolor}
\usepackage{units}
\usepackage{textcomp}
\usepackage{amsmath}
\usepackage{graphicx}
\usepackage{setspace}
\usepackage[authoryear]{natbib}
\PassOptionsToPackage{normalem}{ulem}
\usepackage{ulem}
\makeatletter

\providecommand{\tabularnewline}{\\}
\providecolor{lyxadded}{rgb}{0,0,1}
\providecolor{lyxdeleted}{rgb}{1,0,0}

\DeclareRobustCommand{\lyxdeleted}[3]{{\color{lyxdeleted}\lyxsout{#3}}}
\DeclareRobustCommand{\lyxsout}[1]{\ifx\\#1\else\sout{#1}\fi}

\newcommand{\lyxaddress}[1]{
\par {\raggedright #1
\vspace{1.4em}
\noindent\par}
}

\renewcommand{\lyxdeleted}[3]{ }

\usepackage{pdfcomment}
\usepackage{environ}
\usepackage[left,modulo]{lineno}

\newcommand{\mut}{^\text{m}}

\newcommand{\um}{u\mut}
\newcommand{\fm}{f\mut}

\makeatother

\begin{document}

\title{Coevolutionary patterns caused by prey selection}

\maketitle
\begin{center}\author{Sabrina B. L. Araujo$^{1,2,*}$, Marcelo Eduardo Borges$^{2,3}$, Francisco W. von Hartenthal$^{3}$ , Leonardo R. Jorge$^{4,5}$ , Thomas M. Lewinsohn$^{4}$, Paulo R. Guimarães Jr.$^{6}$ and Minus van Baalen$^{7}$}
\end{center}

\lyxaddress{$^{1}$Departamento de Física, Universidade Federal do Paraná, 81531-980,
Curitiba, Paraná, Brazil.}

\lyxaddress{$^{2}$Laboratório de Ecologia e Evolução de Interações, Biological
Interactions, Universidade Federal do Paraná, P.O. Box 19073, PR 81531-980,
Curitiba, Brazil.}

\lyxaddress{$^{3}$ Pós-Graduação em Ecologia e Conservação, Setor de Ciências
Biológicas, Caixa Postal 19031, Curitiba, PR, 81531-990, Brazil.}

\lyxaddress{$^{4}$Departamento de Biologia Animal, Instituto de Biologia, UNICAMP,
Campinas, Brazil.}

\lyxaddress{$^{5}$Department of Ecology, Institute of Entomology, Biology Centre,
Czech Academy of Sciences, České Budějovice, Czech Republic.}

\lyxaddress{$^{6}$Departamento de Ecologia, Instituto de Biociências, USP, São
Paulo, Brazil.}

\lyxaddress{$^{7}$IBENS CNRS UMR8197, ENS, Paris, France.}

\lyxaddress{{*}Corresponding author: araujosbl@gmail.com}

\newpage

\section*{Abstract}

Many theoretical models have been formulated to better understand
the coevolutionary patterns that emerge from antagonistic interactions.
These models usually assume that the attacks by the exploiters are
random, so the effect of victim selection by exploiters on coevolutionary
patterns remains unexplored. Here we analytically studied the payoff
for predators and prey under coevolution assuming that every individual
predator can attack only a small number of prey any given time, considering
two scenarios: (i) predation occurs at random; (ii) predators select
prey according to phenotype matching. We also develop an individual
based model to verify the robustness of our analytical prediction.
We show that both scenarios result in well known similar coevolutionary
patterns if population sizes are sufficiently high: symmetrical coevolutionary
branching and symmetrical coevolutionary cycling (Red Queen dynamics).
However, for small population sizes, prey selection can cause unexpected
coevolutionary patterns. One is the breaking of symmetry of the coevolutionary
pattern, where the phenotypes evolve towards one of two evolutionarily
stable patterns. As population size increases, the phenotypes oscillate
between these two values in a novel form of Red Queen dynamics, the
episodic reversal between the two stable patterns. Thus, prey selection
causes prey phenotypes to evolve towards more extreme values, which
reduces the fitness of both predators and prey, increasing the likelihood
of extinction.

\section*{Keywords}

coevolution, predator-prey, matching phenotype, predator strategy,
victims preference, prey selection

\section{Introduction}

Antagonistic interactions are an ubiquitous phenomenon in nature,
in which one species gains resources at the expense of another. Victim-exploiter
relationships encompass a range of interspecific interaction modes
such as plant-herbivore, prey-predator and host-pathogen interactions.
These interactions can exert reciprocal selective pressures resulting
in genetic changes in populations; what is defined as coevolution
(\citealt{Janzen1980}). In the last decades, many theories have been
formulated to explain and better understand how interacting species
affect the evolution of others (e.g. \citet{Thompson2005} and \citet{Abrams2000}).

One of the most influential hypotheses in coevolution is the Red Queen
Dynamics (\citealt{vanValen1973}), which proposes that species must
constantly adapt as a response to the unceasing adaptations of the
organisms with which it interacts. As the fitness of the exploited
species is reduced, selection will favor those organisms with a better
capability of defending themselves or evading exploiters, whereas
exploiters will be selected to evolve countermeasures (\citealt{Hochberg1998}).
Many empirical examples and systems that attain such dynamics are
already well known and studied, as between birds and avian brood parasites
(\citealt{Aviles2006,Martin-Galvez2006,Refsnider2010,Noh2018}) and
between herbivorous insects and their host plants (\citealt{ehrlich_butterflies_1964,Chew1977,Nylin2005,Merrill2013}).

Two frequent ingredients present in coevolutionary antagonistic models
are stabilizing and interaction selections. Stabilizing selection
favors an optimum phenotype that the population would evolve towards
in absence of any other selective pressure. The interaction selection
depends on both victim and exploiter phenotypes and, once the interaction
is antagonistic, it does not favor the convergence between victim
and exploiter phenotypes. Although the intensity of the interaction
selection is able to limit the range of phenotypes that an exploiter
can succeed in attacking, it does not limit the phenotype range that
an exploiter will try to attack. It means that the individual behavior
in searching victims is not under selection when no other pressure
aside from stabilizing and interaction selections is considered (\citealt{Matsuda:1985}).
Some models on optimal foraging strategies have already highlighted
the possibility of prey selection (victim choice) may have important
consequences on the population dynamics (\citealt{StephensKrebs:1986,Berec:2000}),
however they do not incorporate evolutionary dynamics. On the other
hand, most models for predator-prey coevolution do not impose any
priority on predators' attacks, which imply that they attack at random
in their predation neighborhood (see e.g, \citealt{Murdoch:1969,May:1974,Hutson:1984,Matsuda:1985,GleesonWilson:1986,Gendron:1987,FryxellLundberg:1994,AbramsKawecki:1998,vanBaalenEtal:2001}).
An exploiter can increase its fitness if it can choose the victim
that maximizes its chances of success instead of interacting randomly.
A classical empirical example is brood parasites, known to coevolve
with their hosts (\citealt{Rothstein1990}). Brood parasites can choose
to parasitize a host when their eggs match their host's egg colour
(\citealt{Resetarits1996,Aviles2006,Refsnider2010,Soler2014}). The
choice of host has consequences for offspring success and therefore
it is subject to strong selection (\citealt{Resetarits1996,Refsnider2010}).
There is also a vast literature on insect oviposition patterns suggesting
that host plant preference and selection has a significant role in
the coevolutionary history of these species (\citealt{Chew1977,jorge_integrated_2014,Nylin2005,Merrill2013}).

The evolutionary patterns predicted by coevolutionary models are a
result of different pressures in which coevolutionary interactions
occur (\citealt{Thompson2005}) Specifically, when the interaction
selection is determined by the similarity of the interacting species’
phenotypes - phenotype matching (\citealt{S.Brown1992,Berenbaum1998,Gomulkiewicz2000,Abrams2000,Gandon2002,Nuismer2006,Calcagno2010,Yoder2010a,Gokhale2013,Andreazzi2017})
- prey phenotype can evolve to values adjacent to the phenotype predators
aim at. When stabilizing selection and non-directional interaction
selection are combined and assume a symmetrical shape (e.g., as a
Gaussian function), the predicted temporal population phenotype distributions
are also symmetrical with respect to the optimum phenotype favored
by stabilizing selection, resulting in\emph{ symmetrical coevolutionary}
\emph{branching }(\citealt{S.Brown1992,Abrams2000,Calcagno2010,Yoder2010a})
or\emph{ symmetrical coevolutionary cycling} (\citealt{Gomulkiewicz2000,Abrams2000,Dieckmann1995}).
In the former, the population phenotypes evolve towards a set of stable
phenotypes symmetrically distributed around the optimum phenotype
favored by stabilizing selection while in the latter the phenotypes
oscillate around it. On the other hand, when the interaction selection
is directional (for example when the outcome of the interaction is
determined by phenotype differences among interacting species), prey
have a preferential evolution pathway imposed by the interaction (\citealt{Abrams2000}).
As a consequence, the symmetry is broken (this is clearly shown by
\citet{Yoder2010a}). Even though there are abundant studies modeling
co-evolutionary patterns, we still lack in understanding how the individual
behavior in victim preference by the exploiters modulates these patterns.

To investigate the effects of prey selection in the coevolutionary
dynamics of antagonistic populations, we approach the evolutionary
predator-prey system where the individual phenotypes related to the
interaction can evolve subject to both interaction (phenotype matching)
and stabilizing selections. These selective pressures are assumed
to have symmetrical shape, which is expected to promote symmetrical
coevolutionary patterns: \emph{symmetrical coevolutionary} \emph{branching}
and \emph{symmetrical coevolutionary} \emph{cycling}. We explore the
coevolutionary outcomes in phenotypic evolution considering two scenarios:
(i) Predation occurs at random; (ii) predators select which prey to
attack among those present in the predator's `attack neighborhood',
according to phenotype matching. We then analyze the individuals'
payoffs and also propose an individual-based model where individual
phenotypes are explicitly modeled and predators have a limited predation
neighborhood. Both approaches agreed that prey selection can promote
an asymmetrical stable pattern. Moreover, simulation outcomes allowed
us to assess the robustness of our findings and also its sensitivity
under different strengths of interaction selection and carrying capacity.

\section{Methods}

\begin{doublespace}
\noindent From now on, we will refer the two trophic level populations
as prey and predators, but they can also stand for herbivores and
food plants, or for parasites or parasitoids and their hosts.
\end{doublespace}

Consider two finite populations of predators and prey. The phenotype
of each individual \emph{i} in a given generation is represented by
a real number, $u_{i}$ ($v_{i}$), where $i$ identifies the individual
and $u$ ($v$) the prey (resp. predator) species. The phenotypes
are heritable but can evolve over generations due to mutation and
selection. All individuals, regardless the trophic level, are submitted
to a stabilizing selection towards the same phenotype value, which
in absence of any other selection would promote convergence of prey
and predator phenotypes. However, the interaction selection, here
modeled as the probability of a predator successfully attacking a
prey, is proportional to their phenotype matching which can prevent
such convergence. Each predator is assumed to be able to interact
only with a small subset of the prey individuals, that corresponds
to the number of individuals in its predation neighborhood. Whatever
the actual spatial distribution of prey and predator populations,
predators have never access to the whole prey population. Typically,
they perceive only a limited subset, the individuals that happen to
be in what we will call the predation neighborhood. What follows is
based on the insight that predators will be presented with an actually
quite limited choice of potential prey. Often, this may be just one
individual but often also a predator may actually have to make a choice.
As we will show, such choices may have profound consequences. Here
we aim to better understand the consequence of the evolutionary dynamics
for two attack strategies: (\emph{i}) \emph{Without prey selection:}
the predator attacks prey in its predation neighborhood at random;
(\emph{ii}) \emph{With prey selection}: the predator directs its attacks
according to phenotype matching, prioritizing the prey that maximize
the chances of successful attack. These alternatives may seem irrelevant
if the probability of a successful attack for a given interaction
is constant. In fact, if the predator has only one prey in its predation
neighborhood, or the predator attacks all available prey or even if
these prey have the same phenotype, both attack strategies are equivalent.
However, if prey phenotypes are different and the predator prefers
to attack some prey in its predation neighborhood over others, the
evolutionary phenotype dynamics can be different. 

Below we first define the stabilizing and interaction selections,
then we write the total fitness for two specific scenarios and then
we assess the evolutionary stable solutions analytically. Finally,
we impose specific rules ( e.g. reproduction, mutation, population
growth) in a computational simulation to test our analytical results
in a specific setup.

\subsection{Stabilizing and interaction selections}

Stabilizing selection favors an optimum phenotype that the population
would evolve towards in absence of any other selection. Here it is
modeled as:

\begin{eqnarray}
h\left(s_{i}\right) & = & \exp\left[-\gamma_{u}(s_{i}-s_{opt})^{2}\right]\:,\label{eq:external}
\end{eqnarray}

\begin{doublespace}
\noindent where $s_{i}\in\{u_{i},v_{i}\}$, is the phenotype of individual
$i$ ( prey if $s=u$, or predator if $s=v$) , $\gamma_{s}$ the
strength and $s_{opt}$ the static optimum phenotype imposed on population
by stabilizing selection. Here we will assume $u_{opt}=v_{opt},$
which promotes convergence of predator and prey phenotypes.
\end{doublespace}

The interaction selection here is described by the success of an attack,
and it only depends on the matching of predator and prey phenotypes,
according to

\begin{doublespace}
\begin{equation}
f\left(u_{i},v_{i}\right)=\exp\left[-\alpha(u_{i}-v_{i})^{2}\right],\label{fattack}
\end{equation}
where $\alpha$ is the interaction strength; a positive value that
defines the specificity of the predator requirements as a function
of the matching between predator and prey phenotypes (the higher $\alpha$,
the lower the probability of successful attack for imperfect matching).
Predation success increases with the similarity of phenotypes, thus
promoting divergence of prey and predator phenotypes. The evolutionary
dynamics mediated by these two selection pressures was well studied
by \citet{S.Brown1992}, where they show that the evolutionary stable
solutions are always symmetric in relation to the static optimum phenotype
imposed by the stabilizing selection: predators evolve to a monomorphic
population where their phenotype is equal to $v_{opt}$ while prey
evolve to a dimorphic population where the two phenotypes are equally
far from the $v_{opt}$ (they also assumed $u_{opt}=v_{opt})$; or
both predator and prey evolve to dimorphic populations, but kipping
the absolute distance (symmetry) of the phenotypes in relation to
the static optimum phenotype imposed by the stabilizing selection.
\end{doublespace}

\subsection{Analytical approach}

In order to understand the effect of predation strategy, we assume
that the energy required for the predator to produce offspring can
be obtained from one unique prey. As a consequence, predators stop
attacking after their first success or after they have tried to attack
all prey in their predation neighborhood. It means that the predation
strategy (order of the attacks: with or without prey selection) can
affect the dynamics.

An ideal method to compare the two predation strategies would be to
write the mean field equations and analyze their Evolutionary Stable
Solutions (\citealt{S.Brown1992,Taylon1978}). However, it turned
out complicated when we tried it. Thus, we analyze only the payoff:
the probability of winning the interaction and producing an offspring.
Besides, we considered two simple cases: (a) Predator and prey are
monomorphic populations; (b) Predators are monomorphic while prey
are dimorphic. For these we first identified prey and predator phenotypes
that mutually maximize the payoffs, which implies phenotypes that
populations would evolve to if not allowed diversification (keep monomorphic
or dimorphic, depending on the cases). Next, we fixed prey and predator
phenotypes on these values and considered that a new prey phenotype
appears. In accordance with custom in adaptive dynamics theory we
will call the dominant phenotype as the ``resident'', and the individual
with the new phenotype as the ``mutant'' (\citealt{Metz1992}). We
then analyzed the payoff of the mutant to check the evolutionary stability
of (a) and (b).
\begin{doublespace}

\subsubsection*{Monomorphic populations: Stability of the asymmetrical pattern}
\end{doublespace}

\begin{doublespace}
We first consider a scenario in which one monomorphic prey and one
monomorphic predator populations interact. The interest of considering
only monomorphic prey is that we do not need to impose any selection
strategy: both models, with and without prey selection, become equivalent.
We will use this as a starting point to assess the consequences of
prey selection.
\end{doublespace}

When a prey and a predator interact, their respective payoffs (the
probability of winning the interaction and producing an offspring)
can be written as

\begin{doublespace}
\begin{eqnarray}
P_{v}(u,v) & = & fh_{v}\:,\label{eq:firness1ux1v_v}\\
P_{u}\left(u,v\right) & = & (1-f)h_{u}\:,\label{eq:fitness1ux1v_u}
\end{eqnarray}
where $f$ is the fitness benefit for the predator due to the probability
of a successful attack (Eq. \ref{fattack}), and $h_{s}=\exp\left[-\gamma s^{2}\right]$
is the stabilizing selection with static optimum phenotype $u_{opt}=v_{opt}=0$,
Eq. (\ref{eq:external}). Observe that the term $(1-f)$ means the
probability of a prey to survive an attack. If we look for the phenotypes
($u$ and $v$) that simultaneously maximize these payoffs, the following
conditions must be satisfied: $\frac{\partial Ps}{\partial s}=0$
and $\left.\frac{\partial^{2}P{}_{s}}{\partial s^{2}}\right|_{u^{\star};v^{\star}}<0$.
The first condition is equivalent to 
\begin{eqnarray*}
\left\{ \gamma+\alpha\left(u-v\right)-\gamma\exp\left[-\alpha(u-v)^{2}\right]\right\} \exp\left[-\alpha(u-v)^{2}-\gamma u^{2}\right] & = & 0\:,\\
\left[\alpha\left(u-v\right)-\gamma v\right]\exp\left[-\alpha(u-v)^{2}-\gamma u^{2}\right] & = & 0\:.
\end{eqnarray*}
One potential solution is $u=v=0$. However, as the second derivative
of the prey payoff is positive, 
\[
\left.\frac{\partial^{2}P{}_{u}}{\partial u^{2}}\right|_{u^{*}=0;v^{\star}=0}=2\alpha>0,
\]
$u=0$ is not a maximum. Assuming $\left|u-v\right|<1$, and deriving
the Taylor expansion to the second order of $(u-v)$ we have $\exp\left[-\alpha(u-v)^{2}\right]\approx1+\alpha\left(u-v\right)^{2}$
, obtaining the approximate pair of solutions

\begin{eqnarray}
u^{*}=\pm\frac{\sqrt{\alpha+\gamma}}{\gamma}; &  & v^{*}=\pm\frac{\alpha}{\gamma\sqrt{\alpha+\gamma}}.\label{eq:Monomorphic_solution}
\end{eqnarray}
Observe that $u^{*}-v^{*}=\pm1/(\alpha+\gamma)$ so that the condition
$\left|u^{*}-v^{*}\right|<1$ is satisfied if $\alpha+\gamma>1$.
The second criteria, $\left.\frac{\partial^{2}P_{s}}{\partial s^{2}}\right|_{u^{\star};v^{\star}}<0$,
 shows that this asymmetrical scenario is stable. It means, that
if we allow only monomorphic populations, their phenotypes would evolve
to $(u^{*},v^{*})=\left(\frac{\sqrt{\alpha+\gamma}}{\gamma},\frac{\alpha}{\gamma\sqrt{\alpha+\gamma}}\right)$,
or $(u^{*},v^{*})=\left(-\frac{\sqrt{\alpha+\gamma}}{\gamma},-\frac{\alpha}{\gamma\sqrt{\alpha+\gamma}}\right).$
(Figure S1 in supplementary material shows the plots for $P_{v}(u^{*},v)$
and $P_{u}(u,v^{*})$). We call it asymmetrical because the stable
phenotypes are below or above the optimum phenotype imposed by the
stabilizing selection. At this point we are not able to argue that
this asymmetrical pattern is evolutionarily stable, that is, if the
phenotype evolutionary pattern would keep asymmetric if we allow mutations.
Below we evaluate the evolutionary stability of this asymmetry.
\end{doublespace}

Now consider the case where predators have two prey individuals in
their neighborhoods. If the prey population is monomorphic not much
changes except that the predators have now two chances to attack a
prey. Suppose prey and predator phenotypes have evolved to $u^{*}$
and $v^{*}$, respectively, and let a new prey phenotype $\um$ appear.
We call the dominant phenotype $u^{*}$ the ``resident'', and $\um$
the ``mutant'' (\citealt{Metz1992}). We consider now that there are
two prey in a given predation neighborhood; a resident and a mutant.
The fate of the mutant will depend on the predator's strategy. If
the predator attacks randomly, the mutant has a $50\%$ chance of
being attacked first. However if the predator is selective it depends
whether the mutant is the most preferred prey or not. That is, if
$\um$ is closer to $v^{*}$ than $u^{*}$ the predator will attack
it; on the other hand, if $\um$ is further away than $u^{*}$ the
predator will consider it as the second option. Thus, the fitness
of the mutant depends not only on its own phenotype, but also on that
of its conspecifics as well as the predators' strategies. If predators
attack at random, the payoff of the mutant prey is thus given by

\begin{doublespace}
\begin{eqnarray}
P_{\text{{random}}}^{\textrm{m}} & (u^{*},\um,v^{*})= & \frac{1}{2}\left[f+\left(1-f\right)\left(1-\fm\right)+\left(1-\fm\right)\right]h\mut,\label{eq:fitness1ux1v_invader_random}
\end{eqnarray}
where $\fm=\exp\left[-\alpha(\um-v^{*})^{2}\right]$ and $h\mut=\exp\left[-\gamma\left(\um\right)\text{\texttwosuperior}\right]$
are respectively the probability of a successful attack and the stabilizing
selection on the mutant. The term $\nicefrac{1}{2}$ refers to the
probability of one of the any two prey being attacked first. If the
resident is attacked first, the mutant can escape from successful
predation with probability $f+\left(1-f\right)\left(1-\fm\right)$,
which means that the resident can be successfully attacked ($f$)
or, if it is not ($1-f$), the attack on the mutant is not successful
($1-\fm$). If the mutant is attacked first, its chance of escaping
is $\left(1-\fm\right)$, the next to last term in Eq.(\ref{eq:fitness1ux1v_invader_random}).
\end{doublespace}

If predators attack selectively, however, the mutant prey's payoff
depends in a discontinuous fashion on its strategy, depending on whether
it is preferred by the predator or not,

\begin{doublespace}
\begin{equation}
P_{\text{{prey\:selection}}}^{\textrm{m}}(u^{*},\um,v^{*})=\begin{cases}
\left(1-\fm\right)h\mut & \text{{if}}\:|\um-v^{*}|\leq|u^{*}-v^{*}|\\
\left[f+(1-f)(1-\fm)\right]h\mut & \text{{if}}\:|\um-v^{*}|>|u^{*}-v^{*}|
\end{cases},\label{eq:fitness1ux1v_invader_random_preference}
\end{equation}
so if the mutant is less preferred it only risks an attack if the
predator failed the attack on the preferred prey. Payoff functions
for the predator and the resident prey are shown at supplementary
material (Figure S2), but here we only show mutant payoff in order
to demonstrate how its fitness varies with predator strategy.
\end{doublespace}

\begin{figure}
\includegraphics[width=7cm]{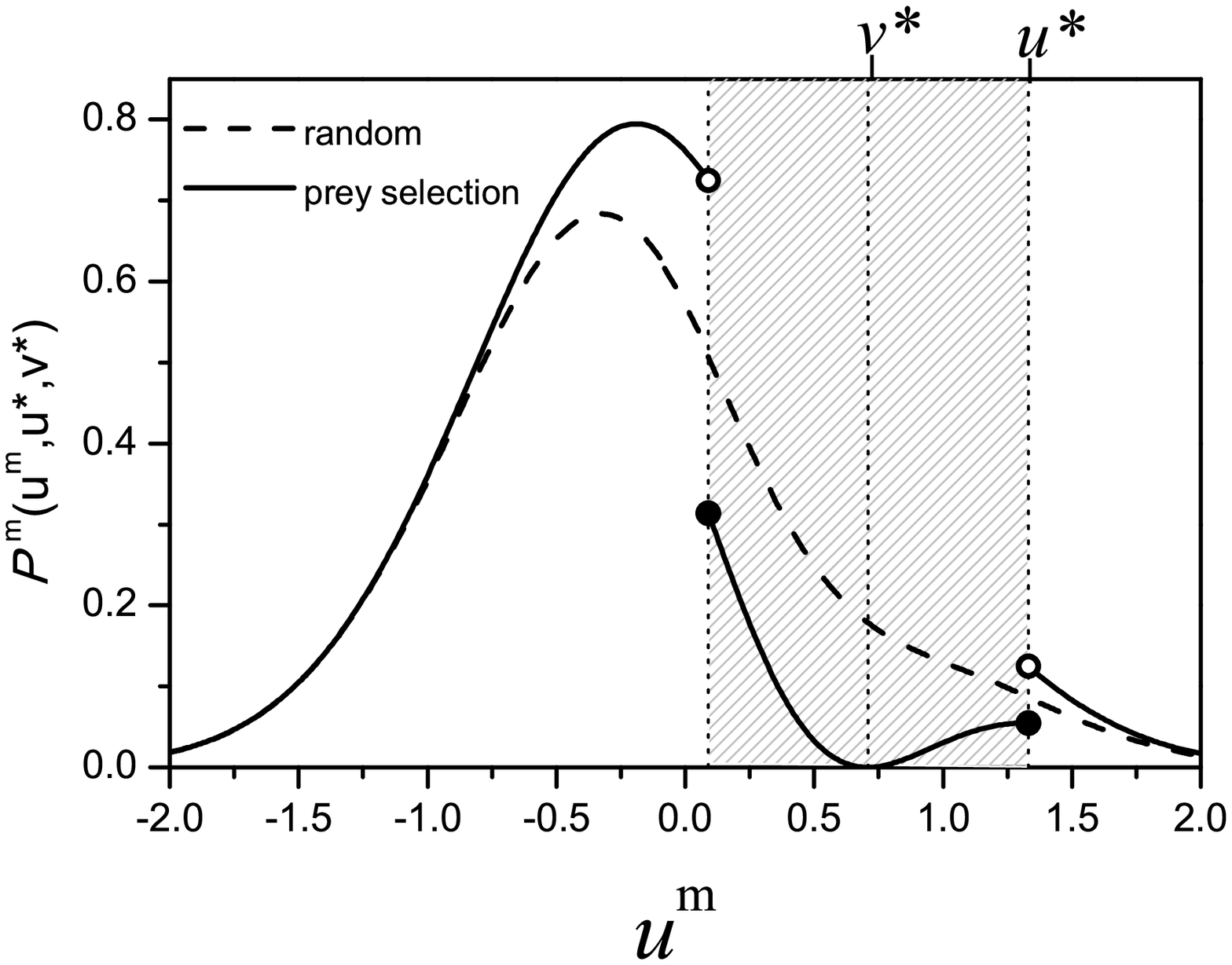}\hspace{1cm}\includegraphics[width=7cm]{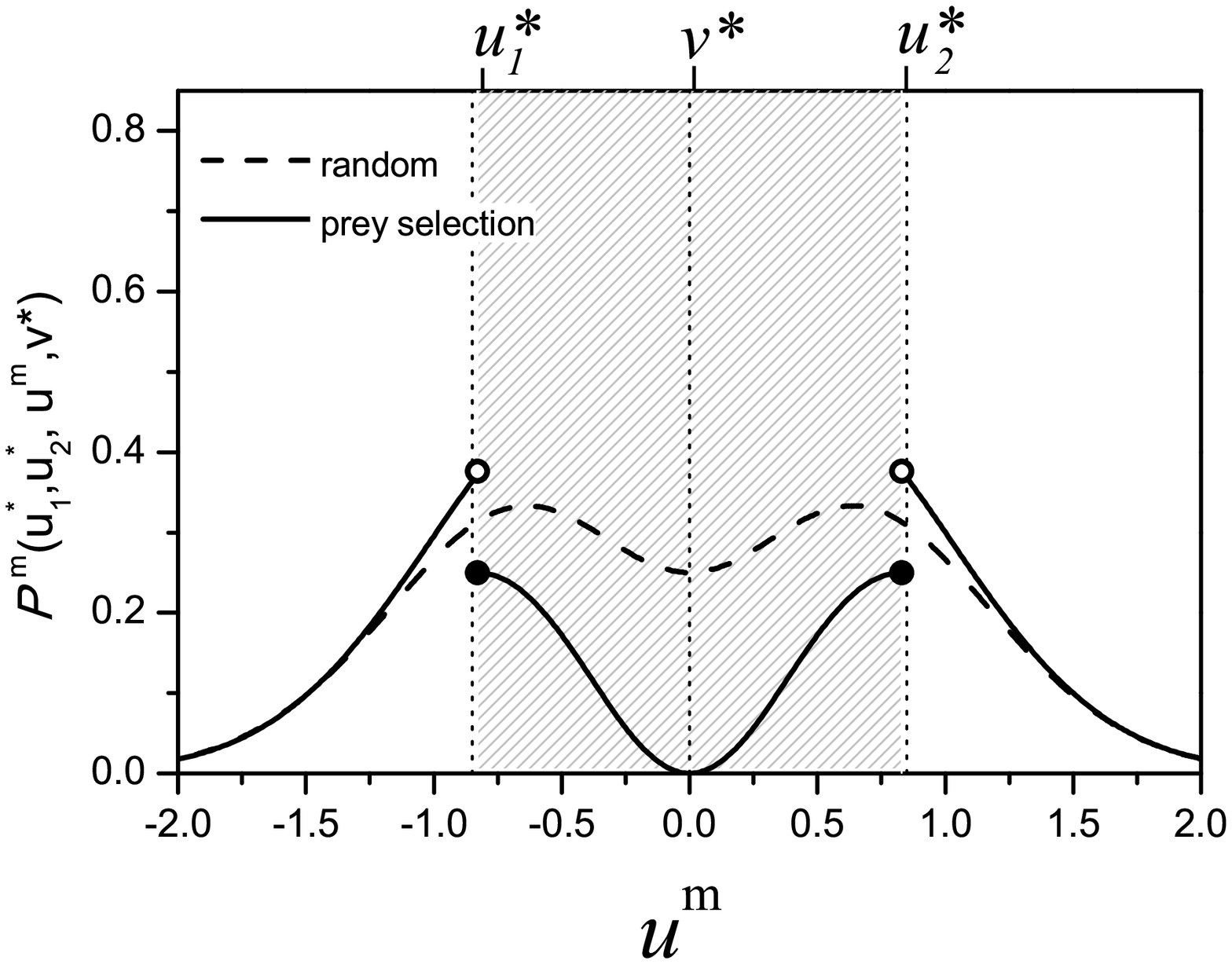}

\caption{Payoffs for mutant prey under the two predation attack strategies:
without prey selection (random; dashed line) and with prey selection
(solid line). \textbf{Left}: prey and predator resident populations
are monomorphic populations and their phenotypes are set at their
stable values, Eq. (\ref{eq:Monomorphic_solution}), represented by
$u^{*}$and $v^{*}$, respectively. The gray area shows the predator's
preferred phenotype range, where $|\um-v^{*}|<|u^{*}-v^{*}|$. If
we consider a mutant prey whose phenotype is similar to the resident
one, the evolutionary outcome will be totally dependent on the predation
strategy: The dashed line, obtained from $P_{\text{{random}}}^{\textrm{m}}(u^{*},\um,v^{*})$
in Eq.(\ref{eq:fitness1ux1v_invader_random}), shows that if predation
occurs randomly the mutant phenotype will evolve towards the maximum
payoff and then will break the asymmetrical pattern; The solid line,
obtained from $P_{\text{{prey\:selection}}}^{\textrm{m}}(u^{*},\um,v^{*})$
in Eq.(\ref{eq:fitness1ux1v_invader_random_preference}), has two
maxima (one when $\um=u^{*}$and a higher one around $-0.4$), however
there is a valley when $u^{m}=v^{*}$ and the higher peak cannot be
attained with small mutation steps. Then, only large mutations can
break the asymmetrical pattern. \textbf{Right}: prey resident population
is dimorphic while predator population is monomorphic and their phenotypes
are setting at their stable values, Eq. (\ref{eq:Dimorphic_solution}),
$u_{1}^{*}$, $u_{2}^{*}$, and $v^{*}$,respectively. The gray area
shows the predator's preferred phenotype range. Both payoff functions,
obtained from $P_{\text{{random}}}^{\textrm{m}}(u_{1}^{*},u_{2}^{*},\um,v^{*})$
in Eq (\ref{eq:1vx2u Invader_Random}) and $P_{\text{{prey\:selection}}}^{\textrm{m}}(u_{1}^{*},u_{2}^{*},\um,v^{*})$
in Eq. (\ref{eq:1vx2u invader_preference}), have two maxima which
means that the symmetrical prey branching can be a stable pattern
regardless the predator strategy. However, when the predation occurs
with preference there is a deeper valley minimizing the flux from
one mode to the other. Parameters $\gamma=\alpha=1$. \label{fig:Fitness-}}
\end{figure}

The consequences of predators preferring certain prey is that in a
homogeneous resident prey population, mutant prey that are slightly
farther away from the `focus' of the predator are better protected.
In the situation depicted in the left graph of Figure (\ref{fig:Fitness-}),
the phenotype combination $(v^{*},u^{*})$ is unilaterally optimal
(i.e., a Nash equilibrium) and thus an ESS candidate. To the prey,
it does not pay to reduce costs by reducing $\um$, as they would
immediately be attacked preferentially, to the predators it does not
pay to focus on larger prey because of the balancing selection. There
is a higher optimum for the prey, but this cannot be reached by small
mutation steps without crossing the adaptive valley. Only when $u^{*}$
and $v^{*}$ are close, mutations may arise that `jump' the valley
and thus break the asymmetric pattern. In larger populations or mutation
rates, the symmetric pattern is likely to result, but in smaller,
stochastic populations, the asymmetric pattern may just switch to
the other side.
\begin{doublespace}

\subsubsection*{Dimorphic prey population: Symmetrical branching stability}
\end{doublespace}

Assume now a situation where there is a monomorphic predator population,
with phenotype $v$, and a dimorphic prey resident population, with
phenotypes $u_{1}$ and $u_{2}$. As before, we will look for the
stable solution when there is only one prey ( $u_{1}$ or $u_{2}$)
in a given predation neighborhood, and then we do not need to impose
any selection strategy. The probabilities that prey and predators
win the interaction and produce an offspring can be written as

\begin{doublespace}
\begin{eqnarray}
P_{v}(u_{1},u_{2},v) & = & \frac{1}{2}\left\{ f_{1}+f_{2}\right\} h_{v}\:,\label{eq:1vx2u_noinvider_v}\\
P_{u_{1}}(u_{1},v) & = & (1-f_{1})h_{u_{1}}\:,\nonumber \\
P_{u_{2}}(u_{2,}v) & = & (1-f_{2})h_{u2}\:,\label{eq:eq:1vx2u_noinvider_u}
\end{eqnarray}
where $f_{s}=\exp\left[-\alpha(u_{s}-v^{*})^{2}\right]$ and the term
$\nicefrac{1}{2}$ assume that prey populations occur with equal abundance.
Analogous to the previous procedure (however without any approximation),
the first and second derivatives allow us to calculate a possible
solution

\begin{equation}
u_{1}^{*}=\pm\sqrt{\frac{1}{\alpha}\ln\left(\frac{\alpha+\gamma}{\gamma}\right)};\:u_{2}^{*}=-u_{1}^{*};\:v^{*}=0\:,\label{eq:Dimorphic_solution}
\end{equation}
that is evolutionarily stable (Fig. S3 in the supplementary material)
when $(\alpha+\gamma)>2\alpha\ln\left(\frac{\alpha+\gamma}{\gamma}\right)$
(if we assume $\gamma=1$ , for example, then the condition is $\alpha<1.37$).
(For greater values of $\alpha$, $v^{*}=0$ becomes a minimum point
and two symmetrical maximum points emerges for $P_{v}(u_{1}^{*},u_{2}^{*},v)$,
leading to divergence in the predator population (\citealt{Vincent1989})).
\end{doublespace}

Regarding the stable symmetric pattern, where the phenotypes evolve
to $u_{1}^{*}$, $u_{2}^{*}$ and $v^{*}$, we approached the effect
of another prey (mutant) in the predation neighborhood, whose phenotype
is $\um$. As before, consider now that a new mutant appears in a
given predation neighborhood. Then, there are two prey in this predation
neighborhood; a resident ($u_{1}^{*}$ or $u_{2}^{*}$ with equal
probability) and a mutant ($\um$). The probability that the mutant
prey is not attacked and produces offspring depends on their phenotypes
and on the predator strategy. If predators attack at random,

\begin{doublespace}
\begin{eqnarray}
P_{\text{{random}}}^{\textrm{m}} & (u_{1}^{*},u_{2}^{*},\um_{i},v^{*})= & \frac{1}{2}\left\{ \frac{1}{2}\left[f_{1}+f_{2}+\left(2-f_{1}-f_{2}\right)\left(1-\fm\right)\right]+\left(1-\fm\right)\right\} h\mut\:,\label{eq:1vx2u Invader_Random}
\end{eqnarray}
otherwise,

\begin{equation}
P_{\text{{prey\:selection}}}^{\textrm{m}}(u_{1}^{*},u_{2}^{*},\um,v^{*})=\begin{cases}
\left(1-f_{i}\right)h\mut & if\:|\um-v^{*}|\leq|u^{*}-v^{*}|\\
\frac{1}{2}\left[f_{1}+f_{2}+(2-f_{1}-f_{2})(1-\um)\right]h\mut & if\:|\um-v^{*}|>|u^{*}-v^{*}|
\end{cases}\:.\label{eq:1vx2u invader_preference}
\end{equation}

Both payoff functions have two maximum points (see the right graph
of Figure \ref{fig:Fitness-}), which means that the symmetrical prey
branching can be a stable pattern regardless of the predator strategy.
However, if the predator attacks with prey selection, the two optimum
phenotypes for the mutant coincide with the phenotypes of the residents
while if the predation occurs randomly they are closer; which means
that prey selection can impose a more extreme phenotype when compared
to random attack. Moreover, if the predator attacks selectively, the
two prey lineages are well separated, $P_{\text{{pref}}}^{\textrm{m}}(u_{1}^{*},u_{2}^{*},\um,v^{*})=0$,
and then the flux from one lineage to the other occurs only by large
mutations (see Figure \ref{fig:Fitness-}). If one of the lineages
goes extinct, selective predation will probably drive the phenotypic
adaptations towards the asymmetrical pattern, while for random predation
large mutations are not required for the extinct lineage to reemerge.
Payoff functions for the predator and the resident prey are shown
at supplementary material (Figure S4).
\end{doublespace}

\subsection{Individual Based Model}

An Individual Based Model (IBM) was built in order to verify the situations
where the asymmetrical pattern is observed. We propose a dynamics
where individuals are submitted to both interaction and stabilizing
selections and the offspring inherit their parent's phenotype plus
mutation, which allows coevolution of the phenotypes. We model the
two predation strategies: with and without prey selection. Some variations
of the model assumptions were made to test the robustness of our conclusions
and are presented at the \textquotedbl{}Robustness\textquotedbl{}
subsection below.

We defined a finite population composed by $M_{u}$ prey and $M_{v}$
predators. Space is not modeled explicitly and for simplicity, the
model considers synchronized events and a discrete life cycle; in
a given generation, the members of both populations first interact
and then reproduce to form the next generation. \textcolor{black}{Although
space is not explicitly modeled, we assume that each predator $i$
can attack only a subset of $n_{i}$ individuals of the prey population
chosen at random, which resembles the limitation of prey in a predator
foraging area (predation neighborhood). We assume that the average
number of prey in a predation neighborhood is $M_{u}/M_{v}$ with
variation among predators. }As the number of predators increases,
the predation neighborhoods diminish, which models the scenario where
predators compete for prey.\textcolor{black}{{} So, we distribute $M_{u}$
prey (with }replacement\textcolor{black}{) over the $M_{v}$ predators
(with} replacement\textcolor{black}{). As consequence,} $n_{i}$ follows
a binomial distribution where\textcolor{black}{{} a prey has a chance
of $1/M_{v}$ of being in each predation neighborhood in each one
of $M_{u}$ events (see details at supplementary material).} We have
also explored the situation where $n_{i}$ is a fixed parameter over
individuals and over time (see \textquotedbl{}Robustness\textquotedbl{}
section), and we show that the conclusions highlighted here do not
depend on the way this is chosen. In both approaches, one prey can
be in more than one predation neighborhood or in no predation neighborhood
at all.

\begin{doublespace}
After setting the prey for each predator's neighborhood, all predators
that have at least one prey in their predation neighborhood will attack.
The sequence of predator attacks is set at random, but once one predator
is chosen, it attacks until its first success or until it has tried
to attack all prey in its predation neighborhood. The probability
of a successful attack is determined by the interaction pressure,
Eq. (\ref{fattack}). If an attack is successful, the prey dies and
the predator will have an opportunity to produce an offspring. A prey
can only be successfully attacked once, and if it occurs, the prey
is removed from the other predation neighborhoods into which it pertains.
This process resembles the interaction between hosts and parasitoids
where parasitoids attempt to deposit eggs inside their hosts, but
hosts may defend themselves, e.g., by encapsulating the parasitoid
egg (\citealt{Bartlett1966}), whereas the successful development
of the parasitoid results in the death of the prey. We have also considered
the situation where the victim does not die, but its fitness is reduced
due to the consumer. Again, our highlighted conclusions are robust
under this model variation.

Only surviving prey, $M'_{u}$, and successful predators, $M'_{v}$,
will reproduce, and for simplicity we consider asexual reproduction
only. The probability of an individual having one offspring depends
on how its phenotype is well adapted to the stabilizing selection,
Eq. (\ref{eq:external}). Moreover, in order to have an upper bound
in prey population size, it was considered intraspecific competition
pressure, resulting in the following probability of a survivor prey
having one offspring:

\begin{eqnarray}
g(u_{i}) & = & \frac{1}{1+\rho(M'_{u}-1)}h(u_{i}),\label{eq:competition}
\end{eqnarray}
where $\rho$ is a parameter that controls the strength of competition
and $h(u_{i})$ is the stabilizing selection, Eq. (\ref{eq:external}).
When there is a unique prey in the whole system there is no competition
and $g(u_{1})=h(u_{1})$, while as the population grows $g(u_{i})$
goes to zero. Predator population size is indirectly limited by the
size of the prey population, so the probability of each predator $i$,
among $M'_{v}$ feeding predators, to have one offspring is given
by $g(v_{i})=h(v_{i})$, Eq. (\ref{eq:external}). After all interactions,
number of offspring which will recompose the next generation is computed
as:
\end{doublespace}

\begin{equation}
M''_{s}=F_{s}\sum_{i=1}^{M'_{X}}g(s_{i}),\label{eq:offspring}
\end{equation}
where $F_{s}$ is a parameter that represents the average number of
offspring per prey (if $s=u$, or predator if $s=v)$ individual under
higher fitness, $h(s_{i}=s_{opt})=1$. The parameter $\rho$, in Eq.
(\ref{eq:competition}), can be written in terms of prey carrying
capacity ($K$) and $F_{u}$. For that, considering the situation
of higher fitness for all $M'_{u}$ prey, the number of offspring
in the next generation would be:

\[
M''_{u}=\frac{F_{u}M'_{u}}{1+\rho(M'_{u}-1)}.
\]
When the population achieves the carrying capacity (K), we will have
$M'_{u}=M''_{u}=K$ , which allows us to write:

\[
\rho=\frac{F_{u}-1}{K-1}.
\]
During the simulation we first calculated the number of offspring,
Eq. (\ref{eq:offspring}) and then associated each offspring to a
parent with probability described in Eq. (\ref{eq:competition}) (
for prey or $g(v_{i})=h(v_{i})$, Eq. (\ref{eq:external}) for predators).
An offspring possesses the same phenotype as its parent plus a normally
distributed mutational variation $\delta_{s}$, with

\begin{doublespace}
\begin{equation}
P(\delta_{s})=\frac{1}{\sigma_{s}\sqrt{2\pi}}\exp\left[-\frac{\delta_{s}^{2}}{2\sigma_{s}^{2}}\right],\label{mutation}
\end{equation}
where $\sigma_{s}$ is the standard deviation. The new generation
replaces the previous generation ($M_{s}=M"_{s}$), a new set of interactions
occurs, the surviving prey and fed predators have offspring, and the
cycle restarts. A list of all parameters involved in the model is
shown in Table (\ref{parameters}).

{\small{}}
\begin{table}
{\small{}\caption{List of all parameters involved in the model, their values utilized
in the simulations and a short description of their meaning}
\label{parameters}}{\small\par}

{\small{}}%
\begin{tabular}{lcl}
\hline 
{\small{}parameter} & {\small{}value} & {\small{}short meaning}\tabularnewline
\hline 
\hline 
{\small{}$\alpha$} & {\small{}\{1, 2, 4, 6, 8, 10, 20, 40, 100, 1000\}} & {\small{}Intensity of selection imposed by the interaction}\tabularnewline
{\small{}$\gamma_{u};\gamma_{v}$} & {\small{}1} & {\small{}Intensity of stabilizing selection pressure}\tabularnewline
{\small{}$u_{opt};\:v_{opt}$} & {\small{}0} & {\small{}Optimum phenotype imposed by the stabilizing selection}\tabularnewline
{\small{}$F_{u};\:F_{v}$} & {\small{}\{2, 4, 6, 8, 10, 12\}} & {\small{}Fecundity}\tabularnewline
{\small{}$K$} & {\small{}\{500, 2000, 5000, 10000, 50000\}} & {\small{}Prey carrying capacity}\tabularnewline
{\small{}$\sigma_{u};\:\sigma_{v}$} & {\small{}\{0.01, 0.02\}} & {\small{}Standard deviation that defines the mutation amplitude}\tabularnewline
\hline 
\end{tabular}{\small\par}
\end{table}
{\small\par}
\end{doublespace}

\subsubsection{Scenarios}

We investigated the coevolutionary patterns (phenotype distributions
over the generations) and population sizes for both models (\emph{with
prey selection} and \emph{without prey selection}) considering different
parameter combinations (see Table\ref{parameters}) but we fixed the
strength of stabilizing selection and the optimum phenotype favored
in absence of the interaction ($\gamma_{u}=\gamma_{v}=1;\;u_{opt}=v_{opt}=0$).
For all simulations the initial condition corresponded to one thousand
individuals of each trophic level and phenotype values were equal
to the optimum phenotype imposed by the stabilizing selection, plus
a normally distributed variation (as in Eq.\ref{mutation}). Each
simulation was iterated over ten thousand generations.

\subsubsection{Robustness\label{subsec:Robustness}}

\begin{doublespace}
To assess the robustness of our results we analyzed some modifications
to the model assumptions. The modifications and results are outlined
below (and detailed in the supplementary material).
\end{doublespace}

We modeled the antagonistic interaction by promoting a fitness benefit
for the exploiter and a fitness prejudice for the victim instead of
death due to interaction. The benefit and prejudice are controlled
by two independent parameters, which allows to analyze the effect
of different impacts on each trophic level. For simplicity, the population
size of each trophic level remained constant, then the contribution
of each individual to the next generation population was proportional
to its individual fitness. The number of victim in an exploiter's
attack neighborhood was a fixed parameter and the interaction occurs
only with one individual. As in the original model, we investigate
two scenarios: (i) attack occurs at random; (ii) exploiters can select
a victim according to phenotype matching. We ran simulations for 400
combinations of parameters for each scenario (with and without prey
selection): we qualitatively predict the same coevolutionary patterns
observed originally, including the exclusivity of the asymmetrical
pattern for the second scenario (See Figures S7 and S8, in the supplementary
material).

\section{Simulation results}

\begin{doublespace}
\begin{figure}
\includegraphics[width=17cm]{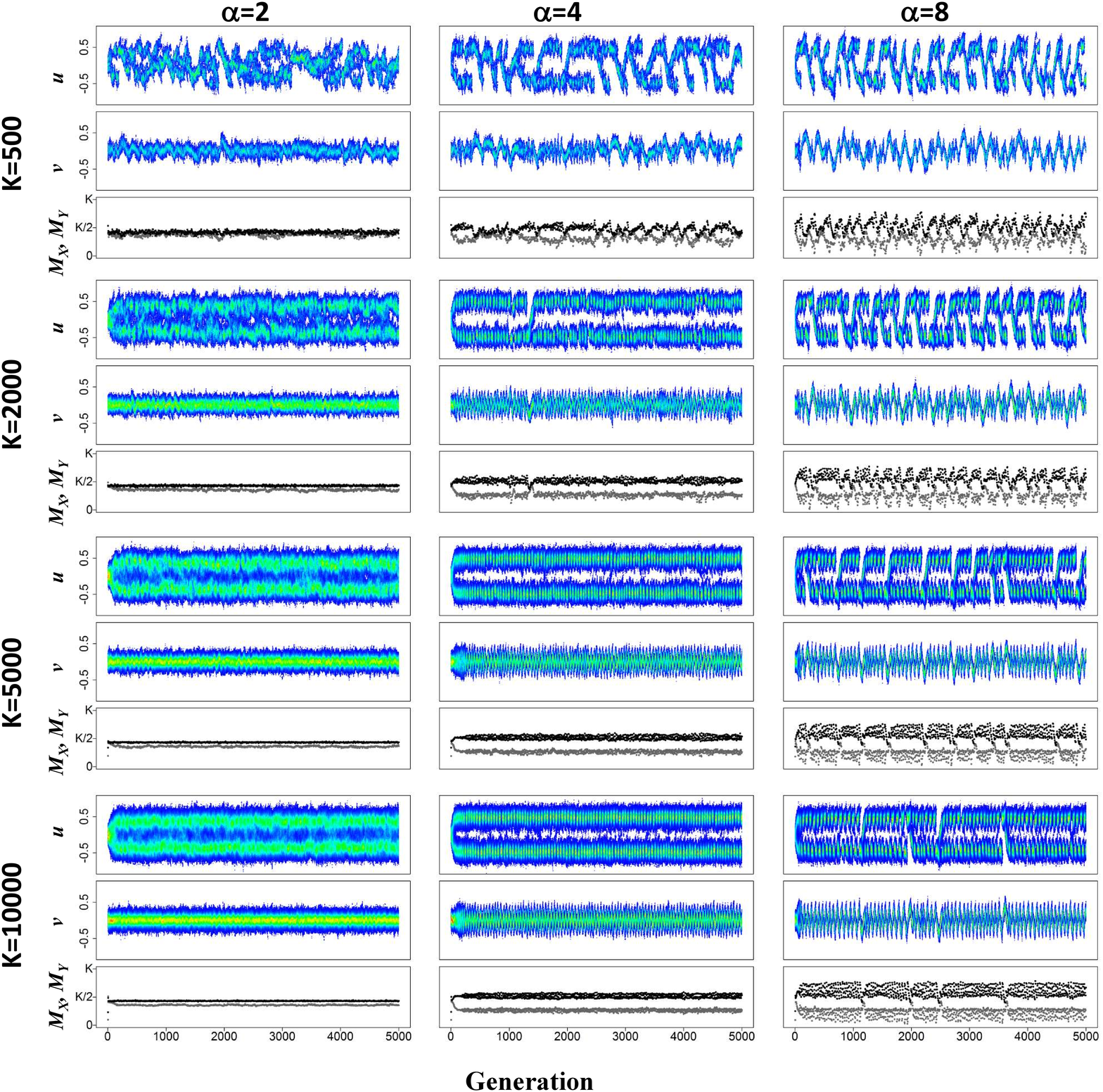}\caption{Coevolutionary temporal patterns for the model\emph{ without prey
selection} and with different intensities of interaction strength
($\alpha$, in columns) and carrying capacity ($K$, in rows). The
first two graphs of each set of three sequential graphs show prey
($u$) and predator ($v$) phenotype temporal evolution, respectively.
The color scale is proportional to the number of individuals with
a given phenotype (from blue to red meaning low to high frequencies).
The bottom graph shows the prey (black) and predator (gray) population
size over time. Only the last three values of $K$ correspond to the
same parameter values present in Figure \ref{fig:Pref}. Parameters:
$\gamma=1,$ $u_{opt}=v_{opt}=0$, $F_{u}=8$, $F_{v}=2$, $\sigma_{u}=\sigma_{v}=0.02$.
\label{fig:NoPref}}
\end{figure}

\begin{figure}
\includegraphics[width=17cm]{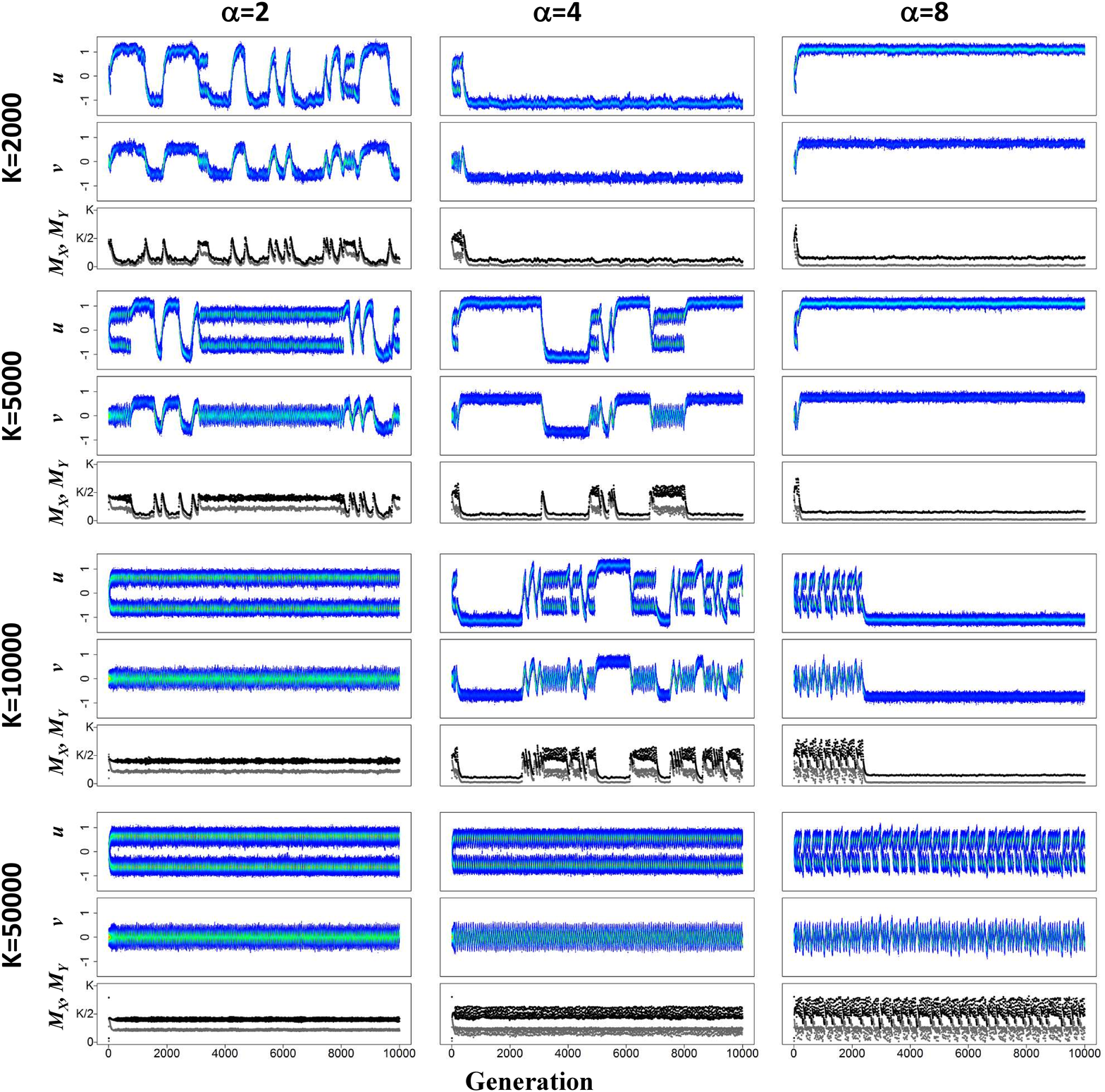}

\caption{Coevolutionary temporal patterns for the model\emph{ with prey selection}
and considering different intensities of interaction ($\alpha$, in
columns) and carrying capacity ($K$, in rows). The first two graphs
of each set of three sequential graphs show prey ($u$) and predator
($v$) phenotype temporal evolution, respectively. The color scale
is proportional to the number of individuals with a given phenotype
(from blue to red meaning low to high frequencies). The bottom graph
shows the prey (black) and predator (gray) population size over time.
Only the first three values of $K$ correspond to the same parameter
values present in Figure \ref{fig:NoPref}. Parameters: $\gamma=1,$
$u_{opt}=v_{opt}=0$, $F_{u}=8$, $F_{v}=2$, $\sigma_{u}=\sigma_{v}=0.02$.\label{fig:Pref}}
\end{figure}

\end{doublespace}

Both models were sensitive to the interaction strength ($\alpha$)
and the carrying capacity ($K$) (Figures \ref{fig:NoPref} and \ref{fig:Pref}).
Higher values of $\alpha$ lead to a more intense selection in the
predator population whose phenotypes better matches their prey phenotypes.
From the prey point of view, the pressure results in the differentiation
of their phenotypes from the predator population. The condition of
attacks \emph{with prey selection}, Figure \ref{fig:Pref}, has similar
effects to increasing interaction selection, but the evolutionary
pattern observed under this strategy cannot always be recovered by
increasing the interaction strength ($\alpha$) in the model \emph{without
prey selection}, Figure \ref{fig:NoPref} (detailed below). The carrying
capacity is also an important parameter for the evolutionary outcomes
since smaller populations are more vulnerable to demographic stochasticity
and extinction (Figures \ref{fig:NoPref} and \ref{fig:Pref}). Once
prey selection increases the pressure on prey, the minimum value of
$K$ to observe predator and prey coexistence is higher when compared
to the attacks without prey selection (Observe that the minimum value
of $K$ differs between Figures \ref{fig:NoPref} and \ref{fig:Pref}).

We qualitatively classify three different patterns for the population
phenotype distribution across the generations: \emph{symmetrical coevolutionary
oscillations, symmetrical coevolutionary} \emph{branching }and \emph{asymmetrical
pattern}. In \emph{symmetrical coevolutionary oscillations}, prey
and predator phenotypes oscillate around the respective optimum values
in absence of the interaction (as an example see Figure \ref{fig:NoPref}
when $K=500$ and $\alpha=8$ and Figure \ref{fig:Pref} when $K=2000$
and $\alpha=2$). In this scenario, neither population reaches an
equilibrium phenotype distribution. Occasionally a bifurcation in
the distribution of the prey phenotype appears that last for a few
generations but that will eventually disappear again when one of the
branches goes extinct. In the\emph{ symmetrical coevolutionary} \emph{branching}
pattern, prey phenotypes bifurcate between two lineages with phenotype
values symmetric in relation to the optimum phenotype imposed by the
stabilizing selection, while the predator phenotypes assume values
between both lineages and around $v_{opt}=0$ (as an example see Figures.\ref{fig:NoPref}
when $K=10000$ and $\alpha=\{2,4\}$ and \ref{fig:Pref} when $K=50000$
and $\alpha=\{2,4\}$). \textit{\emph{Finally, the}}\emph{ asymmetrical
pattern} was observed exclusively in the model with prey selection.
Both prey and predator phenotypes evolve to values above or below
(some simulations can result above while another simulation with same
parameters can result below) the static optimum phenotype imposed
by stabilizing selection but without tallying up; predator phenotypes
stay between the prey phenotype and the optimum phenotype imposed
by the stabilizing selection (see Figure \ref{fig:Pref} when $(K;\alpha)=\{(2000;4),$$(2000;8),$
$(5000;8),$ $(10000;8)\}$). In agreement with our analytical approach,
predators that can selectively attack a preferred prey can lock the
prey phenotype in one of the two stable phenotypes. Any prey mutant
that tries to cross towards the other stable phenotype becomes the
preferred prey, which minimizes its reproductive success. When the
attack is random (without prey selection) that mutant prey will never
be preferentially attacked. We have also tested if the asymmetry would
emerge by increasing the interaction strength when the attack is random
($\alpha$=\{10, 20, 40, 100, 1000\}, Figure S5), but it resulted
in prey extinction or high frequency phenotype oscillations. We have
also confirmed this result under variation of model assumptions, see
the Robustness section above.

\begin{figure}
\includegraphics[height=6cm]{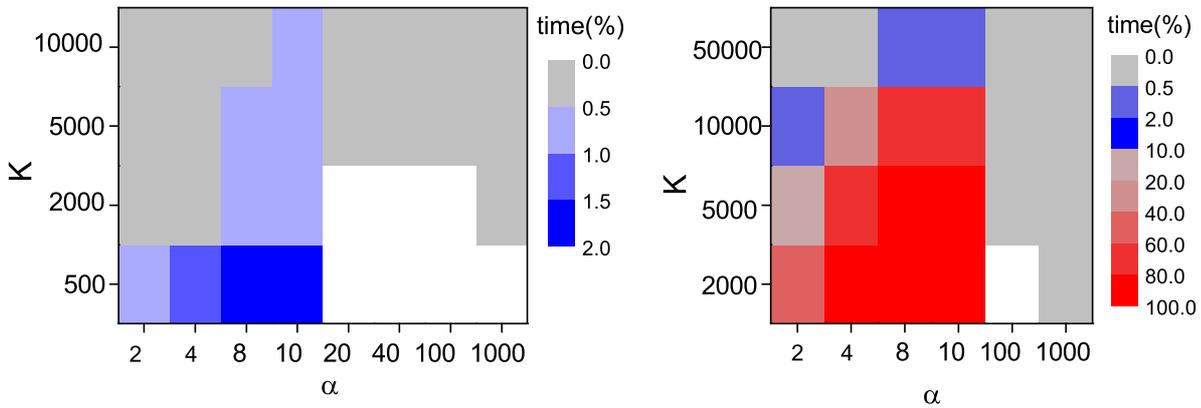}\caption{Average of the longest time in asymmetrical pattern as function of
the interaction strength $(\alpha)$ and carrying capacity ($K$)
for the model \emph{without prey selection}(left) and \emph{with prey
selection }(right). For each parameter combination was analyzed the
longest time in asymmetry for 7000 generations (from generation 3,000
to 10,000). The average was calculated over 10 replicates. The graphs
show the percentage (from 0 to 100\% of analyzed time in asymmetry).
White area means extinction of at least one of the populations for
all replicates. Observe that the graphs have different color scales;
the model without prey selection does not maintain the asymmetry for
more than 2\% of analyzed time, which corresponds to about 125 generations,
while the model with prey selection the asymmetry can be maintained
for all analyzed time.}
\label{Fig-Average}
\end{figure}

We also calculated the maximum time that the prey phenotypes stay
in asymmetry. For that, from generation 3,000 to 10,000 we calculated
the maximum time that the average prey phenotype did not cross zero.
We then calculated the average time over 10 replicates (Figure \ref{Fig-Average}
and Figure S6). Only in the dynamics with preference some parameters
combinations resulted in 7,000 generations (100\% of analyzed time)
in asymmetry for all repetitions. The maximum time that we could observe
asymmetry when prey selection is not considered was about 125 generations
(for other parameter combinations we could observe almost 300 generations,
see Figure S6), which is actually the period of cycling phenotypes,
not stable asymmetry. If we fix a $K$ value and increase $\alpha$,
we have an increase of the time in asymmetrical pattern up to a certain
$\alpha$ value; from that value on, populations are extinct or the
phenotypes oscillate in high frequency (it occurs for both models,
see Figure \ref{Fig-Average} for $\alpha=\{100,1000\}$). The high
frequency oscillations probably occur because the standard deviation
of the interaction selection ($\sim\nicefrac{1}{\sqrt{\alpha}}$,
see Eq.\ref{fattack}) approaches the mutation standard deviation
$\sigma$ (Eq.\ref{mutation}). It means a higher probability of a
survivor prey having a descendant out of the predator phenotype requirements.

\begin{doublespace}
If we consider a given interaction strength $\alpha>0$ and observe
the patterns as a function of $K$, we first have extinction of predators
or of both populations (not shown in Figures \ref{fig:NoPref} and
\ref{fig:Pref}, but in Figure S5 and Figure \ref{Fig-Average}, where
higher values of $\alpha$ are considered). As population densities
increase, predator and prey populations coexist going through the
following patterns: \emph{asymmetrical pattern} (only in the model
with prey selection), \emph{coevolutionary oscillation} and finally
\emph{symmetrical branching} (as an example see Figure \ref{fig:Pref}
when $\alpha=4$). In the transitions between these patterns we may
have combinations of them, for instance with populations that oscillate,
bifurcate and have lineages that go extinct and emerge again (see
Figure \ref{fig:NoPref} when $K=500$ and $\alpha=4$ and Figure
\ref{fig:Pref} when $K=10000$ and $\alpha=4$, for example). A similar
sequence of coevolutionary patterns occurs if we consider a given
value $K$ and decrease $\alpha$.

In addition to the asymmetrical pattern, prey selection produces more
extreme evolutionary dynamics compared to the random predation model:
the phenotypes evolve to values further from the optimum set by stabilizing
selection (even if we compare high $\alpha$ in the model \emph{without
prey selection }with low $\alpha$ in the model \emph{with prey selection}),
decreasing the fitness of both populations. As a consequence, extinction
is more likely to happen when predators attack with prey selection
under low prey carrying capacity. For example, when $K=500$ either
predators or both predator and prey populations become extinct for
any value of $\alpha$, while in the model without prey selection
we observe coevolutionary oscillation (see Figure \ref{fig:NoPref}
when $K=500$). Coevolutionary oscillation also appears in the model
with prey selection for higher values of $K,$ resulting in oscillations
with higher amplitude and periods (compare Figure \ref{fig:NoPref}
when $K=500$ and $\alpha=2$ to Figure \ref{fig:Pref} when $K=2000$
and $\alpha=2$).
\end{doublespace}

\begin{figure}
\includegraphics[width=15cm]{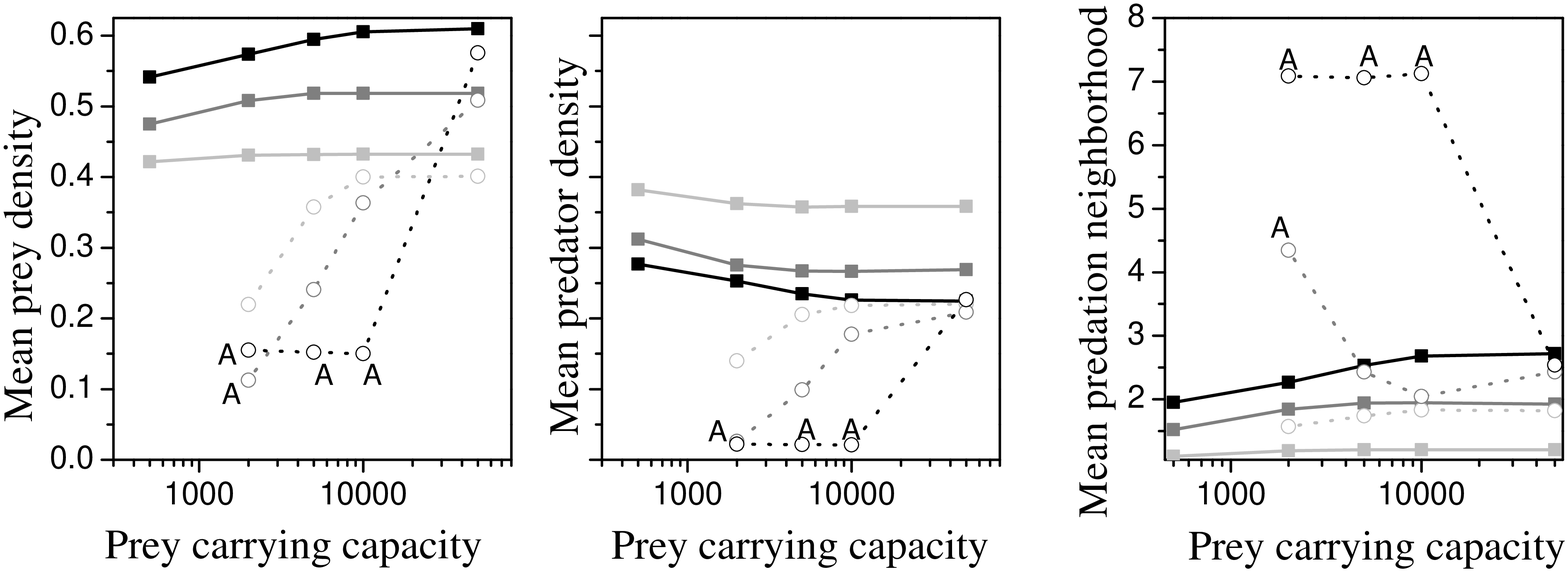}\caption{From left to right: Prey and predator population densities, $\left\langle M_{u}\right\rangle /K$
and $\left\langle M_{v}\right\rangle /K$, and mean size of the predation
neighborhood, $\left\langle n\right\rangle =\left\langle M_{u}/M_{v}\right\rangle $,
as function of prey carrying capacity, $K$. The filled squares refer
to the scenario where predators attack at random (without prey selection)
while the open circles to the scenario with prey selection (the first
point, $K=500$, is not plotted for the case with prey selection because
predator and prey do not coexist). These averages were computed considering
the last $7000$ of $10000$ generations. The colors, light gray,
gray and black correspond to $\alpha$ equal to $2$, $4$ and $8,$
respectively. The letter \textquotedbl{}A\textquotedbl{} highlight
the parameter combinations where the asymmetrical pattern occurs.
\label{fig:density}}
\end{figure}

Population size fluctuations arise in particular under non-equilibrium
dynamics of phenotype evolution (coevolutionary oscillation and in
the transition between patterns) while a more constant population
size was associated with stable phenotype evolutionary patterns (asymmetrical
pattern and symmetrical branching) (Figure \ref{fig:NoPref} and \ref{fig:Pref}).
Although asymmetrical pattern leads to a constant population size,
predator population size was lowest in these cases, making it more
vulnerable to extinction (by increasing the interaction strength or
imposing a perturbation). We also computed the mean size of the predation
neighborhood, $\left\langle n\right\rangle =\left\langle M_{u}/M_{v}\right\rangle $
and mean relative population densities, $\left\langle M_{u}\right\rangle /K$
and $\left\langle M_{v}\right\rangle /K$, calculated from the last
7000 generations of each simulation (Figure \ref{fig:density}). Both
population densities tend to converge for a given interaction strength
($\alpha$), as the carrying capacity increases. Prey populations
remain sensitive to the intensity of interaction regardless of the
predators' strategy, whereas predator populations are sensitive only
if they attack at random; if predators attack preferentially certain
prey, their density seems to converge to the same point independently
of $\alpha$. Consequently, the predation neighborhood increases with
the intensity of the interaction. For high values of $K$, prey selection
leads to a slightly larger mean predation neighborhood. For low values
of $K$, where the asymmetrical pattern occurs, both population densities
are reduced, but the predator population diminishes more drastically.
In compensation, the predation neighborhood increases so that the
few predators have more prey to choose from (see the highlighted dots
in Figure \ref{fig:density}).

\section{Discussion}

We investigated the coevolutionary dynamics of predator-prey systems
in which predators can choose which prey to attack in their immediate
neighborhood. The interaction selection was considered non-directional;
only the difference between prey and predator phenotype determines
predation success. However, in addition to a situation in which predators
encounter prey only randomly, we also allow them to select prey to
attack from prey present in their predation neighborhood, where they
will preferentially attack prey with the closest match. We compare
the individuals' payoffs and the temporal population phenotype distributions
– coevolutionary patterns – for a number of different parameter combinations
and also under different model assumptions (Robustness section), that
represent different ecological settings. Our results support four
main conclusions described below.

First, trait evolution is sensitive to the intensity of interaction
selection and the populations' carrying capacity. For a given value
of interaction strength, increasing the carrying capacity has the
following effects: at small abundances, we have extinction of predators
or both populations; as population densities increase, they tend to
coexist indefinitely, shifting through the following patterns: \emph{asymmetrical
pattern} (which occurs only in the model with prey selection), \emph{symmetrical
coevolutionary oscillation} and finally \emph{symmetrical branching.}
The term \textquotedbl{}symmetry\textquotedbl{} means that, along
time, phenotypes occupied both regions delimited by the optimum phenotype
with equal amplitude. In the transition between patterns, these are
combined with populations oscillating, bifurcating and forming lineages
that go extinct and then evolve again. A similar sequence of coevolutionary
patterns occurs with a fixed carrying capacity and decreasing interaction
strength. With symmetrical branching, the prey evolves to a bimodal
phenotype distribution, while the predator evolves to an intermediate
phenotype which is equally effective against both prey lineages. The
coevolutionary oscillatory pattern is related to Red Queen evolution
\citep{vanValen1973}, where rare prey phenotypes are more likely
to evade predation and thus phenotype frequencies in prey populations
change continuously, while the predators keep evolving to specialize
on the most common prey types (\citealt{Dieckmann1995}). It should
be noted however, that the oscillatory pattern observed under prey
selection is somewhat different, and results from the episodic reversal
between two stable patterns. In contrast to van Valen's Red Queen
dynamics, this oscillatory pattern strongly depends on population
sizes, as it depends on the frequency of those rare mutations that
manage to cross the adaptive valley driven by predators selectively
attacking preferred prey. It is interesting to note that the simulation
has not shown predator phenotype branching (following the prey branching)
that would emerge, according to Brown and Vincent's study \citeyearpar{S.Brown1992}
and also by our analytical approach, when the interaction strength
increases. Instead, as the interaction strength increases, we observed
\emph{symmetrical coevolutionary oscillation.} We did not explore
the reason of it, but we propose that this pattern still may emerge
outside the parameter space we explored, probably by decreasing the
intensity of stabilizing selection and then allowing for a wider range
of phenotypes. Future studies should be done in order to clarify it.

Both stable and unstable patterns have been predicted by many theoretical
models (\citealt{Gomulkiewicz2000,Abrams2000,Dieckmann1995,Levin1977,Thompson2005,S.Brown1992,Calcagno2010}),
although empirical examples remain sparse. The presence of bimodal
host phenotype distributions have been empirically observed in a natural
\emph{Daphnia} population during a parasite epidemic (\citealt{Duffy2008})
and also in host egg coloration subject to avian brood parasitism
(\citealt{Yang2010}). Empirical evidence of an oscillating pattern
was again found in host egg colorations and their avian brood parasite
by \citet{Spottiswoode2013} and in a herbivorous moth and its host
plant (\citealt{Berenbaum1998}). Frequency-dependent selection between
populations has already been pointed by \citet{Levin1977} as an important
driver in evolutionary history. In agreement with these authors, we
verified that selective pressure in smaller populations can lead to
different evolutionary effects from those expected in larger populations.

Second, both the analytical approach and IBM simulations agree that
prey selection causes prey phenotypes to evolve towards more extreme
values than randomly attacking predators. When the resulting evolutionary
pattern is oscillatory, its amplitude is higher and when prey phenotypes
bifurcate, the stable phenotypes are more extreme. This higher differentiation
between phenotypes in the branching pattern suggests that selective
predation can facilitate speciation. This would be caused if phenotypic
differentiation was followed by reproductive isolation \citep{Dieckmann1999}.
The current model does not allow for this additional evolutionary
step, as reproductive isolation makes no sense under the assumption
of asexual reproduction. Future studies, expanding this model to include
sexual reproduction and reproductive isolation could test the effect
of the resource selection strategy on diversification rates in order
to validate this prediction.

Third, prey selection makes predators more efficient which paradoxically,
reduces at least the predator population size. Other models of exploitative
interactions have found that enhancing a consumer’s efficiency reduces
its population density (\citealt{Peterson1984}). In our model, for
a sufficiently high carrying capacity, the predator strategy has no
effect on prey density, or on the ensuing evolutionary pattern: in
both strategies the system converges to symmetrical branching at high
carrying capacity (although the stable prey phenotypes are more divergent
under selective predation). However, under low carrying capacity,
the predator density decreases when there is prey selection (see Figure
\ref{fig:density}), and as a consequence, the predation neighborhood
increases (since it is proportional to the ratio of prey and predator
population sizes). Predator population size is reduced under selective
versus random predation, probably because selective predation imposes
a higher pressure on prey phenotype, which in turn exerts a higher
pressure on predators. In a first instance, for a given prey phenotype
distribution, prey selection increases predator success, but over
evolutionary time it reduces the predator population benefit compared
to random predation. At low values of carrying capacity, both populations
are more vulnerable to extinction when predation is selective. This
issue is particularly important regarding species that evolve in islands,
in fragmented patches or that naturally occur in low abundance.

Fourth, an asymmetrical pattern can emerge unexpectedly. Usually,
when there is a stabilizing selection that bounds the prey and predator
phenotype distributions around the same optimum value and the interaction
pressure is non-directional (phenotype matching, which promotes bidirectional
axis of vulnerability, as approached by \citet{Abrams2000}), the
expected emerged patterns for the coevolving species are symmetric;
oscillating or bifurcating in two or more lineages (around an optimal
value imposed by the stabilizing selection). For species phenotypes
to coevolve in asymmetry, the expected mechanism would be directional
pressure, phenotype difference, which promotes an unidirectional axis
of vulnerability (\citealt{Abrams2000}). Here we showed that non-directional
pressure (phenotype matching) associated to prey selection can limit
prey phenotypes to an asymmetrical pattern, which can only be broken
if population densities increase. In accordance with the analytical
results, the model variation presented in the supplementary material
showed that the asymmetrical pattern is not unique to the set of assumptions
made in the model presented here at the main text. There, population
density does not vary trough time (extinction is not allowed) but
impacts the fitness of populations. The main common assumptions made
in both model variations were the presence of a predation neighborhood
and the predator satiation (once it successfully attacks a prey, it
stops attacking). These assumptions are the key for what makes the
predator strategy relevant: if an optimum prey is in a predation neighborhood,
its fate is to be attacked if the predator strategy considers prey
preference. However, if the predator attacks randomly, that prey can
go unnoticed. It means that predators that can selectively attack
a preferred prey can lock the prey phenotype in one of the two stable
phenotypes. Any prey mutant that tries to cross towards the other
stable phenotype becomes the preferred prey, which minimizes its reproductive
success. Although, to our knowledge, asymmetrical patterns have not
been documented before in the conditions mentioned above, such patterns
can be observed in absence of stabilizing selection (\citealt{Calcagno2010})
and also in models of specialization on two habitats, where strong
fitness trade-offs can produce specialists in only one habitat (then
asymmetrical) (\citealt{Levins1962,Rueffler2004,Ravigne2009}). Our
analytical model shows the importance of sharp decrease in the probability
of survival of the prey under prey preference, when the phenotype
of the mutant is closer to the predator phenotype than the resident
phenotype. This situation can occur when there is a slight difference
between the prey phenotypes, and the predator prefers the phenotype
that gives it the greatest fitness contribution, even if the difference
is very small. This discontinuity may be not realistic, since in a
predation neighborhood a small phenotypic variation may not be noticed
by the predator. As a consequence, a mutant can no longer `hide' behind
more preferred prey, and the asymmetrical pattern would be broken
more easily. However, we expect that the asymmetrical pattern persists,
but less pronounced. Moreover, we have not explored other interaction
and stabilizing functions that were not Gaussian, (Eqs. (\ref{eq:external})
and (\ref{fattack})). Then, the generalization of our results for
other symmetrical functions remains open. Further study should assess
the generalization of it as well as which level of differentiation
by the predator is necessary to produce asymmetrical patterns.

Our study can shed light on currently observed evolutionary patterns,
and a prime example is the evolution of egg morphology of cuckoo brood
parasites and their hosts. This system presents the two elements we
are highlighting: non-random host selection (parasite individuals
choose the nest where they will lay eggs) and non-directional pressure
(parasite success increases with phenotype matching) (\citealt{Aviles2006,Soler2014}).
The study by \citet{Spottiswoode2013} compared the appearance (color
and patterns) of cuckoo finch eggs with their hosts, the tawny-flanked
prinia (\emph{Prinia subflava}) from the same location in Zambia over
40 years. As mentioned before, they found that egg colors seem to
be locked in an ongoing arms race (\emph{symmetrical coevolutionary
cycling}). However, part of the egg color pattern traits measured
have been changing their mean values, accompanied by decreases in
phenotypic variation, suggesting directional selection (evolving to
an asymmetrical pattern which the authors interpreted as resulting
from some undetected directional pressure which would drive this asymmetrical
evolution). Our model provides an alternative explanation for this
result, which requires no other pressure besides the host selection
by brood parasites according to phenotypic-matching; the observed
pattern emerges directly from this non-random interaction.

In conclusion, our results bolster the conclusion that non-random
interactions can have important ecological and evolutionary consequences.
This ubiquitous behavioral aspect of antagonistic interactions had
so far been ignored in co-evolutionary models, and this study shows
that in addition to the long known ecological consequences prey selection
also has unexpected evolutionary consequences, such as generating
striking asymmetrical (or cyclic) evolutionary outcomes. Our results
are also a reminder to keep considering the ecological context in
evolutionary dynamics, as our results are most pronounced under those
ecological conditions where population sizes are relatively low.

\section*{Acknowledgements}

SBLA received research assistanships from the Conselho Nacional de
Desenvolvimento Científico e Tecnologico (CNPq), LRJ was supported
by Fapesp scholarships (predoctoral grant \#09/54806-0; post-doctoral
grant \#14/16082-9), and TML received CNPq productivity grant \#311800/2015-7.
SBLA, LRJ, TML and PRG thank the São Paulo Advanced School on Ecological
Networks (supported by Fapesp grant \#2010/51395-7) for promoting
the collaboration of this work. MvB received support under the program
430 Investissements d'Avenir launched by the French Government and
implemented by ANR with the 431 references ANR-10-LABX-54 MEMOLIFE
and ANR-11-IDEX-0001-02 PSL Research 432 University.

\bibliographystyle{amnat}
\bibliography{Refs}

\end{document}